\documentclass[%
 reprint,
 amsmath,amssymb,
 aps,
prl,
]{revtex4-1}

\usepackage{graphicx}
\usepackage{dcolumn}
\usepackage{bm}
\usepackage{color}
\usepackage[subrefformat=parens,labelformat=parens]{subcaption}

\DeclareMathOperator{\Tr}{Tr}

\begin{document}

\preprint{APS/123-QED}

\title{Accounting for Errors in Quantum Algorithms via Individual Error Reduction}
  
\author{Matthew Otten}
\author{Stephen Gray}%
 \affiliation{%
 Center for Nanoscale Materials, Argonne National Laboratory, Lemont, Illinois, 60439
 }%

\date{\today}

\begin{abstract}
We discuss a surprisingly simple scheme for accounting (and removal) of
error in 
observables determined from
quantum algorithms. 
A correction to the value of the observable is calculated by first measuring the
observable with all error sources active and subsequently measuring the
observable with each error source removed separately. We apply this scheme to
the variational quantum eigensolver, simulating the calculation of the ground state energy of
equilibrium H$_2$ and LiH in the presence of several noise sources, including
amplitude damping, dephasing, thermal noise, and correlated noise. We show that
this scheme provides a decrease in the needed quality of the qubits by up to two
orders of magnitude. In near-term quantum computers, where full fault-tolerant
error correction is too expensive, this scheme provides a route to
significantly more accurate calculations.

\end{abstract}

\pacs{Valid PACS appear here}
\maketitle


\section{Introduction}
Quantum computing, though in its infancy, is beginning to show promising 
proof-of-principle
calculations, especially in quantum chemistry. Calculations of the
binding energy curve for molecules such as H$_2$~\cite{omalley-prx-2016} and
BeH$_2$~\cite{kandala-nature-2017} have been done using small, noisy quantum
computers. Quantum computing is entering what some are calling the noisy
intermediate-scale quantum era~\cite{preskill-arxiv-2018}. Full fault-tolerant
error correction is still many years away; near-term quantum computers will
have a limited number of qubits, and each qubit will be noisy. Methods that
reduce noise and correct errors without doing full error correction on every
qubit will help extend the range of interesting problems that can be solved in
the near-term. 

In this work, we describe and demonstrate a simple scheme for
reducing the effects of a wide variety of noise sources by removing each source
separately and summing the resulting corrections. These noise sources could be
removed by any process, but we imagine that they are removed by quantum error
correction. Simple quantum error correction schemes have already been shown
in systems including superconducting
circuits~\cite{reed2012realization,kelly2015state} and such systems have been
shown to be below the threshold for even more complicated error
correction schemes~\cite{barends2014superconducting}, such as the surface
code~\cite{fowler2012surface}. Quantum computing architectures are nearing the
quality and size where a single qubit could be error corrected, but
we are far from the realm where every qubit can be corrected. The scheme we
present can make use of this limited error correction, by correcting each qubit
separately. We demonstrate this scheme with the
variational  quantum eigensolver (VQE)~\cite{peruzzo-ncomms-2014,mcclean-njp-2016}, 
by simulating
the calculation of the ground state energy of H$_2$ and LiH. We assume that a
single qubit is error corrected, while the other qubits retain all
of their error. Our results show that it reduces the needed quality of each qubit
drastically; for `chemical accuracy', error rates can be up to two orders of magnitude
larger. We apply this scheme to two wavefunction ansatzes and multiple noise
sources, including amplitude decay, dephasing, thermal noise, and correlated
noise. We stress that the scheme can be used to reduce the environmental error from any
measured observable, not just those used in the VQE algorithm.

\section{Theoretical Methods}

\subsection{Time Evolution and Noise Modeling}\label{t-evol}
Consider a system of $n$ qubits characterized by a time-dependent density matrix
$\rho (t)$.  
These qubits are subject to a sequence of $k$ = 1, 2, ..., $G$ 
gate operations, each described by a unitary transformation $U_k$ that 
corresponds to an instantaeous `jump' on $\rho$:
\begin{equation}
  \rho \rightarrow U_k \rho U_k^\dagger.
\end{equation}
We assume a time $\tau$ lapses between each gate operation, and during
these times $\rho$ evolves under a Lindblad master equation
\begin{equation}\label{master_equation}
\frac{\mathrm{d} \rho}{\mathrm{d} t}
  =  L ( \rho )  = \sum_{i=1}^m L_i ( \rho ), 
\end{equation}
where $m$ represents the number of Lindblad terms (alternatively, the number of
error sources); this can be equal to or an integer multiple of the number of
qubits, $n$, but does not need to be.
In our calculations, we simulate four different types of Lindblad operators,
representing varying noise sources:
amplitude damping ($L^1$), dephasing ($L^2$), thermal ($L^{th}$), and a
correlated noise term ($L^c$):
\begin{equation}\label{lindblad-i}
\begin{split}
  L^1( \rho ) &=  \gamma_1\mathcal{D}[\sigma](\rho ),\\
  L^2( \rho ) &= \gamma_2 \mathcal{D}[\sigma^\dagger \sigma]( \rho ),\\
  L^{th} (\rho) &= \gamma_{th} (n_{th} + 1) \mathcal{D}[\sigma] + \gamma_{th} n_{th} \mathcal{D}[\sigma^\dagger],\\
L^{c} ( \rho ) &= \gamma_c \mathcal{D}[\sigma_1^\dagger\sigma_2] + \gamma_c \mathcal{D}[\sigma_1 \sigma_2^\dagger],
\end{split}
\end{equation}
where $\mathcal{D}[C](\rho ) =  C \rho C^\dagger  -
\frac{1}{2}(C^\dagger C \rho + \rho C^\dagger C)$. 
These Lindblad terms
are applied to each qubit or to various combinations of qubits.
The parameters in Eq.~\eqref{lindblad-i} are
$\gamma_1 = \frac{1}{T_1}$, the decay rate
associated with the amplitude damping noise; $\gamma_2 = \frac{1}{T_2}$, the decay
rate associated with dephasing noise; $\gamma_{th}$, the thermalization rate;
$n_{th}$, the thermal occupation (taken to be 0.5 in this work); and
$\gamma_{c}$, the correlated noise rate.
This is the same formalism we have used in
previous work~\cite{macquarrie-ncomms-2017,otten-pra-2016,otten-prb-2015}. For
many of the results, we assume that the Lindblad terms for a single qubit $i$ have been
removed, $\gamma^{i} << \gamma$. In practice, this could be done via quantum error
correction~\cite{nielsen2010quantum}, or by 
some active engineering  which
greatly reduces the noise rate.
In this work, gates are separated by one
time unit and the error rates are given in inverse time units.

\subsection{Variational Quantum Eigensolver}\label{VQE}
Here, we provide a brief overview of the variational quantum eigensolver
(VQE). VQE solves for an approximate, variational, ground state by
optimizing, using a classical computer, the energy of a parameterized
wavefunction ansatz, $|\psi(\theta)\rangle$, which is evaluated on a quantum
computer. The variational principle ensures that
\begin{equation}
  E_0 \le \frac{\langle  \psi(\theta) | H | \psi(\theta) \rangle}{\langle \psi(\theta) | \psi(\theta) \rangle},
\end{equation}
where $E_0$ is the true ground state energy of the Hamiltonian, $H$. $E$, the
energy of the parameterized wavefunction ansatz, is
evaluated on the quantum computer, and the parameters $\theta$ are optimized
using a classical computer. Classical computing methods such as variational quantum Monte
Carlo~\cite{nightingale-qmc-1998} also make use of the same variational
principle; the hope of a quantum realization is that quantum computers can
efficiently prepare non-trivial states that would be much more difficult to
prepare on a classical computer.
Though methods like quantum phase estimation~\cite{kitaev-arxiv-1995} can give
generally more accurate energies, VQE requires shorter circuits and has a
natural robustness to noise~\cite{sawaya-jctc-2016,peruzzo-ncomms-2014,kandala-nature-2017}. When
using VQE to solve for quantum chemistry problems, as this work does, the second
quantized quantum chemistry Hamiltonian is transformed into a Hamiltonian acting
on qubits using a transformation such as
Jordan-Wigner~\cite{whitfield-molphys-2011}. We use the open source package
OpenFermion~\cite{mcclean-arxiv-2017} to generate the qubit Hamiltonian,
starting first from the quantum chemistry integrals, generated via
Psi4~\cite{parrish-jctc-2017}. We use the unitary coupled cluster singles doubles (UCCSD)
ansatz, which is described in detail in Ref.~\cite{peruzzo-ncomms-2014}. We use
OpenFermion~\cite{mcclean-arxiv-2017} and ProjectQ~\cite{steiger-arxiv-2018} to
generate the circuit for the UCCSD ansatz. We evolve the system using the
high-performance density matrix evolution program QuaC: Open Quantum Systems in
C~\cite{Quac:17}. We optimize the parameters of the wavefunctions using both
Nelder-Mead~\cite{nelder-cj-1965} and COBYLA~\cite{powell-cobyla-1994}.

\section{Results}
Consider an initial density matrix $\rho(0)$ and  
let $\rho(T)$ be the density matrix obtained after $G$ gate 
evaluations and no error removal, i.e. between each gate application
we allow Lindblad time evolution for a time $\tau$ as described
above in order to mimic the effects of environmental
noise on the qubits.  If time $t$ = 0 corresponds to the first gate
application and time $t$ = $T$ to the last gate application, 
then $T$ = $(G-1)\tau$.  
Let $\rho_a(T)$ be the corresponding density
matrix if there were perfect error correction or, equivalently in our formalism,
there was no Lindblad evolution between gate applications.  
Finally, 
consider the density matrices $\rho_i(T)$, with $i$ = 1,2,..., $m$, corresponding
to a calculation where Lindblad superoperator $i$ is removed, but all the
others are still active.  We can define corresponding observables
\begin{equation}\label{observables}
  \begin{split}
    \langle A \rangle &= \Tr ( \rho (T) A),\\
    \langle A_a \rangle &= \Tr ( \rho_a(T) A),\\
    \{\langle A_i \rangle &= \Tr ( \rho_i(T) A), i = 1,m \}.
    \end{split}
\end{equation}
We propose that the observable $\langle A \rangle $ can be corrected to yield a
more accurate value; i.e., one closer to the value with no noise sources,
$\langle A_a\rangle$, according to:
\begin{equation}\label{correction}
\tilde{A} = \langle A \rangle - \sum_{i=1}^m (\langle A \rangle - \langle A_i \rangle).
\end{equation}
Suppose, for example, that $m$ = $n$, the number of qubits, and each
qubit is noisy. 
The strength of Eq.~\eqref{correction} is that, for 
near-term quantum computing without the
possibility of perfect error correction of all $n$ qubits, only $\mathcal{O}(n)$ computations
involving just one qubit being error corrected (to yield the $\langle A_i \rangle$) are
required, along with the original calculation with no error correction to yield
$\langle A \rangle$. Intuitively, the difference $\langle A \rangle -
\langle A_i \rangle$ isolates a subset of the contributions caused by the noise
terms; these are then subtracted away from the noisy expectation value, leaving
a result with much less noise. This cancellation relies on the expectation
value being built up from many measurements. Each expectation value ($\langle  A
\rangle$ and $\langle A_i \rangle$) contains contributions from measurements
with no error, as well as measurements with errors. Our correction scheme
cancels out some of the measurements with error, while
leaving the result with no error, leading to a better calculation of the observable.
The Appendix 
shows that
$\tilde{A}$ approximates $\langle A_a \rangle$
with no errors in first order $\tau$, the time interval between gate
applications, whereas 
the uncorrected result $\langle A \rangle$
approximates $\langle A_a \rangle $ with errors already 
appearing in the
first order of $\tau$.  
Our 
simulations of the application of this scheme to VQE indicate that
Eq.~\eqref{correction} does indeed yield a substantial improvement for real 
algorithms.  Eq.~\eqref{correction} lends itself naturally to use
on a quantum computer, where calculations can be repeated and quantum error
correction offers a natural way to remove errors, but it is not restricted to
just that. Any quantum system which in which an observable can be repeatedly
measured and each noise source can be removed separately can make use of the
scheme to obtain a more accurate result.

In VQE, the measured observable in question is the energy, $E$, of the
wavefunction ansatz. We first optimize the parameters of the wavefunction;
this can be done either with no error correction or potentially error correcting
a single qubit. Once a set of optimal parameters is found, the expectation value
of the energy is evaluated on the quantum computer with no error
correction, and then error correcting each qubit separately. For an $n$ qubit
problem, this involves only an additional $\mathcal{O}(n)$ evaluations of the energy on the
quantum computer with error correction on one of the qubits each time. Once all
of the energies are measured, Eq.~\eqref{correction} 
is used to obtain $\tilde{E}$.

\subsection{H$_2$}

\begin{figure}
  \centering
  \includegraphics[width=0.9\columnwidth]{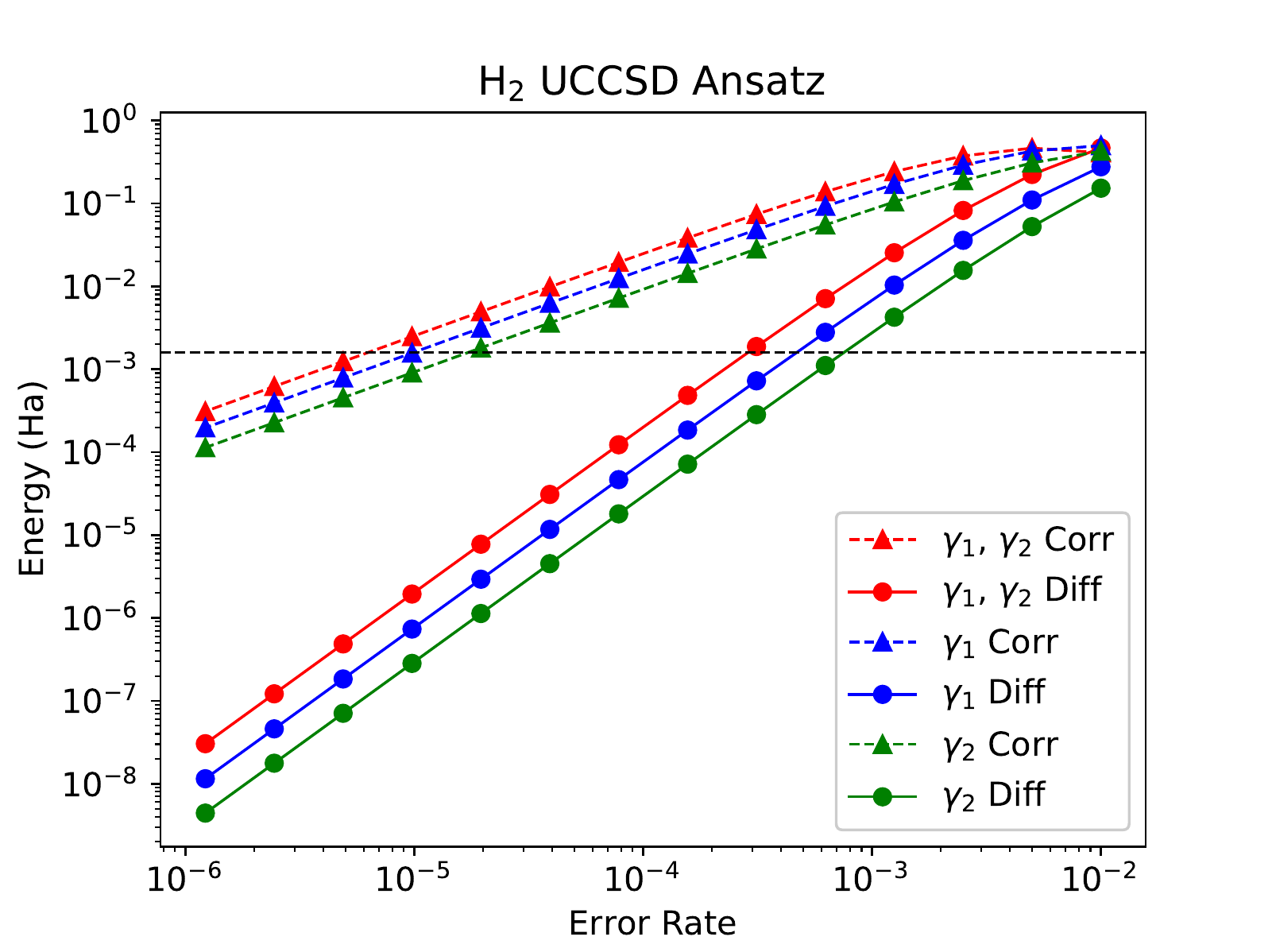}
  \caption{
    Error cancellation for H$_2$ at bond length 0.74 \AA~using the
    unitary coupled cluster singles doubles (UCCSD) ansatz, under amplitude
    damping ($\gamma_1$), dephasing 
    ($\gamma_2$), and both amplitude damping and dephasing ($\gamma_1, \gamma_2$)
    noise sources. The horizontal line represents `chemical
    accuracy', 1.6 mHa. The dashed lines with triangle markers represent the
    amount of correction applied by our scheme. The solid lines with circle
    markers represent the difference between the corrected energy and the energy
    evaluated with no noise. Corrections of up to 70 mHa are applied to get to
    chemical accuracy and error rates can be up to 50$\times$ larger.    
    (Error rates in this and all subsequent figures are in units of the inverse
     time interval between gate applications.)
  } 
\label{h2_example}
\end{figure}

Our first example is the hydrogen dimer, H$_2$, at equilibrium geometry (bond
length 0.74 \AA). We use the sto-3g basis, resulting in a four qubit circuit. We
use the UCCSD ansatz (166 gates) and note that each gate is applied
sequentially with one time unit between each gate; we made no effort to 
apply gates in parallel. The parameters of the wavefunction ansatz
were optimized with noise on every qubit. 
We then sweep through the qubits, removing the noise from one qubit at a time,
simulating the effect of error correction on just that qubit. The final energy
is then calculated by using our correction scheme, Eq.~\eqref{correction}.
We plot the results for typical amplitude damping ($\gamma_1$) and dephasing
($\gamma_2$) type noises in 
figure~\ref{h2_example}, representing three
different environmental regimes.
$\gamma_1 = \gamma_2$ is similar to a 
superconducting qubit quantum computer~\cite{kandala-nature-2017}, whereas the noise on spin~\cite{watson2018programmable} and
trapped ion~\cite{kielpinski2002architecture} quantum computers is dominated by
$\gamma_2$. The dashed lines 
with triangle markers represent the magnitude of the correction used in our
correction scheme, Eq.~\eqref{correction}. The solid lines with circle markers
represent the difference between the corrected energy and the energy if every
qubit were perfectly error corrected. We see that chemical
accuracy (1.6 mHa, represented by the horizontal black line) can be obtained with error rates 
almost two orders of magnitude higher; on average, the error rates can be
45$\times$ larger. To get to chemical accuracy, corrections of 60-70 mHa are
applied. Furthermore, the corrections at all error rates, even the smallest, get
the answer continually closer to the fully error corrected answer. Though it has
not been proved in this paper, this gives evidence that this scheme may be
variational for VQE. The Appendix 
provides results for a different
wavefunction ansatz, one similar to the entangling ansatz of
Ref.~\cite{kandala-nature-2017}. The results are consistent when using this
separate ansatz.
The ordering of the results provides a limited
sensitivity analysis for different quantum computing architectures. Similar to
Ref.~\cite{sawaya-jctc-2016}, we note that VQE is more sensitive to amplitude
damping noise than to dephasing noise.

Even though the wavefunction parameters
were optimized in the presence of noise, the final energy evaluated at the
different parameter sets for the fully error corrected circuit differ very
little. The optimal parameters from the largest error rates only gave a
difference of 1.7mHa compared to the optimal parameters from the error free
optimal parameters, when both were evaluated with no noise; this is much less than the
remaining error (due to noise), even after correction, for the largest error rates. Because of
this, we optimize the parameters once with no noise and use those parameters for
evaluation at all noise rates in the following examples.

\begin{figure}
  \centering
  \includegraphics[width=0.9\columnwidth]{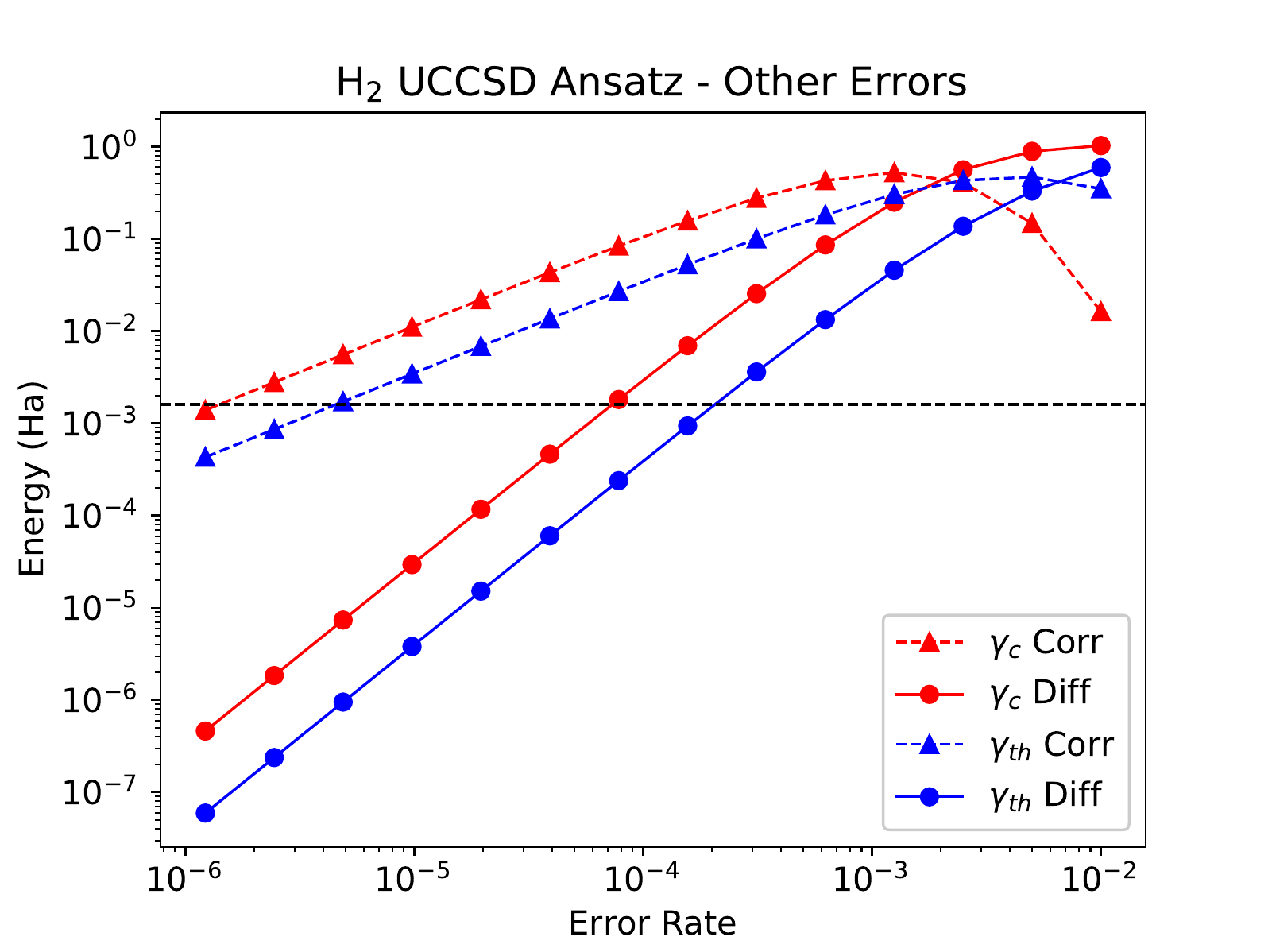}
  \caption{
    Error cancellation for H$_2$ at bond length 0.74 \AA~using the
    unitary coupled cluster singles doubles (UCCSD) ansatz under thermal noise
    ($\gamma_{th}$) and correlated noise ($\gamma_c$). The horizontal line
    represents `chemical
    accuracy', 1.6 mHa. The dashed lines with triangle markers represent the
    amount of correction applied by our scheme. The solid lines with circle
    markers represent the difference between the corrected energy and the energy
    evaluated with no noise. Just as in amplitude damping and dephasing noise
    sources, corrections are as large as 70 mHa and error
    rates can be nearly 50$\times$ larger for chemical accuracy.} 
  \label{h2_other_noise}
\end{figure}

The correction scheme presented in this work is not limited to
environmental noise sources, such as those modeled by $\gamma_1$ and $\gamma_2$,
and removal via error correction.
The scheme is general; systems with any noise source, describable by a Lindblad
superoperator, can benefit, as long as each noise source can
be isolated and removed independently of all other noise sources. To demonstrate
this, we applied a thermal type noise source with rate $\gamma_{th}$ and a
correlated noise source with rate $\gamma_c$ to the H$_2$ UCCSD example. The
results are shown in figure~\ref{h2_other_noise}. The trends are similar to those
for amplitude damping and dephasing; corrections of around 70 mHa bring the
energy to within chemical accuracy at error rates almost 50 times larger than
otherwise needed. Though it might be experimentally difficult, thermal noise
could be reduced by selectively cooling each qubit, one at a time.
The correction scheme applied to our correlated noise
term reveals some subtleties of the method. Our correlated noise Lindblad,
$L^{c} ( \rho ) = \gamma_c \mathcal{D}[\sigma_1^\dagger\sigma_2] + \gamma_c
\mathcal{D}[\sigma_1 \sigma_2^\dagger]$, naturally has terms from two qubits.
When we sweep through the qubits, we remove all terms which involve a single
qubit; this leads to the removal of each $L^{c}$ term {\em twice}, once for each
qubit in each $L^c$. This double counting can be simply corrected by taking half
of the calculated correction from each qubit. Our scheme relies on the fact that
each term is removed once (and only once); as long as the noise sources of interest and their
controlled removal are well understood, the scheme can be applied. If a noise
source is removed twice, accounting for that allows for a good correction. In
our correlated noise term, $L^c$, every term is removed exactly twice and the
calculated correction can be simply halved. Conversely, it is also true that if exactly half
the noise is removed, the correction as if all of the noise is removed can be
calculated by doubling the correction. Though it might be
hard to imagine that something as specific as `half the noise' can be removed,
this idea can be used when the noise is controllable. If the noise is increased
by a controlled, known amount (say, doubled, or even just fractionally
increased), for each qubit separately, the correction scheme can be applied. The
`correction' would be the difference between the inflated noise run and the
normal noise run, scaled by the appropriate factor. This is similar in spirit to
Ref.~\cite{li2017efficient}, where the total noise of the system is artificially
increased and the results are subsequently extrapolated to the zero noise limit.

\subsection{LiH}
\begin{figure}
  \centering
  \includegraphics[width=0.9\columnwidth]{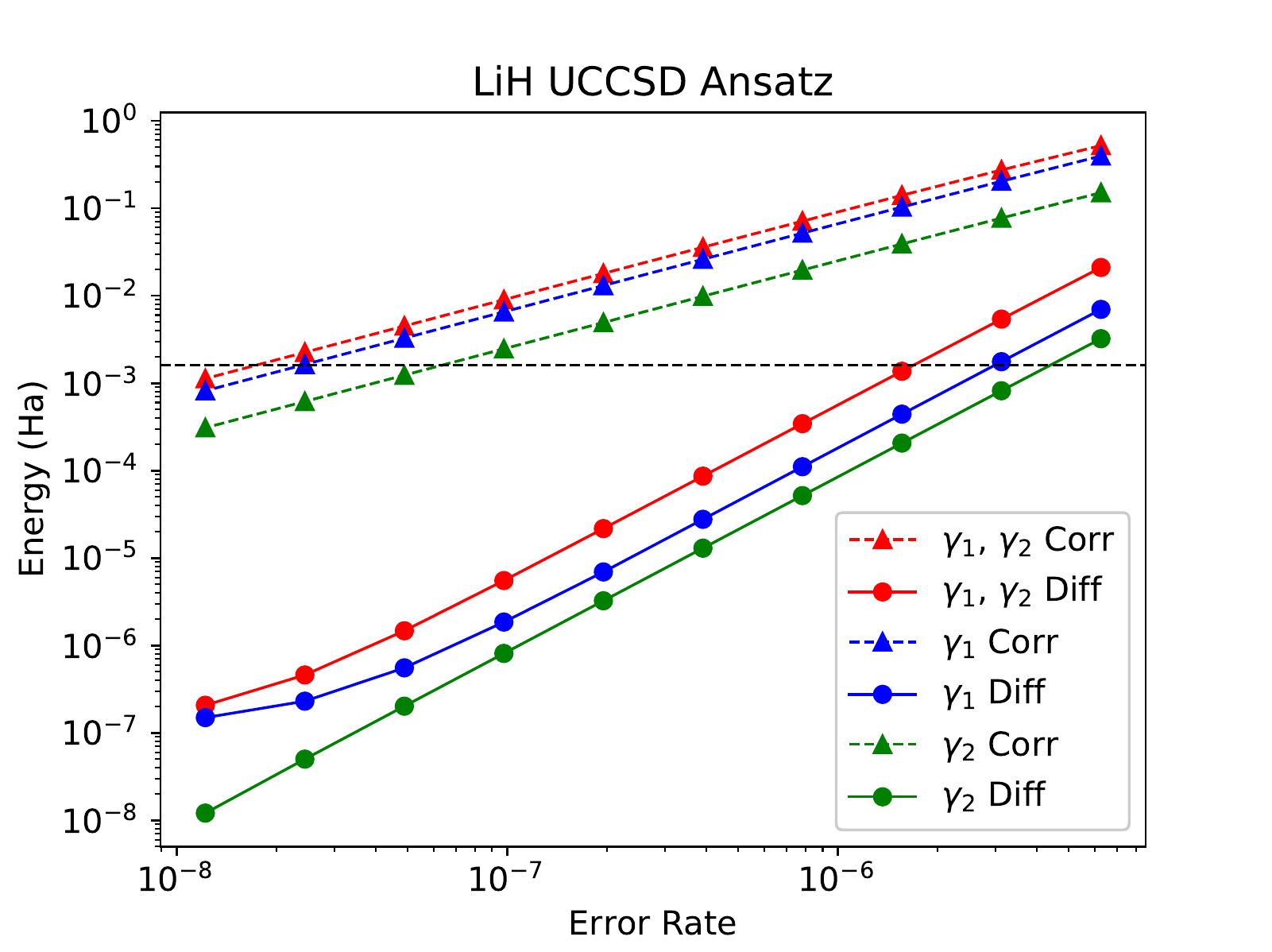}
  \caption{Error cancellation using the unitary coupled cluster singles doubles
    (UCCSD) ansatz for LiH, under amplitude damping ($\gamma_1$), dephasing
    ($\gamma_2$), and both amplitude damping and dephasing ($\gamma_1, \gamma_2$)
    noise sources. The horizontal line represents `chemical accuracy', 1.6 mHa.
    The dashed lines with triangle markers represent the
    amount of correction applied by our scheme. The solid lines with circle
    markers represent the difference between the corrected energy and the energy
    evaluated with no noise. Corrections are between 100-200 mHa and error rates
    can be over 100$\times$ larger for chemical accuracy.}
  \label{lih_example_uccsd}
\end{figure}

We also studied LiH in the sto-3g basis at bond length 1.74 \AA, using 12 qubits, with over 12,000
gates.  We plot the results
for the UCCSD ansatz in figure~\ref{lih_example_uccsd}. Due to the increased
number of qubits and circuit depth, the 
error rates are, overall, much smaller. The correction for LiH is
even more dramatic than for H$_2$. Corrections of 100-200 mHa bring the answer
to within chemical accuracy, and error rates can be over two orders of magnitude
higher, ranging from 68 for with only $\gamma_1$ noise to 128 with only
$\gamma_2$ noise. This example provides confidence that the correction scheme
will work for larger circuits; in fact, it works even better, in this case, than
for the smaller circuits of H$_2$. It is reasonable to believe that this will be
true for ever larger circuits; the number of first order errors increases with
increasing number of qubits, and these are all approximately corrected. Though
the number of second order errors also increases, the second order errors go as
$\gamma^2$ and $\tau^2$, and will generally be small.

\section{Conclusion}
We provide a simple scheme to greatly reduce the error in quantum algorithms and
apply this scheme to simulations of the variational quantum eigensolver. 
Error correcting each qubit, one at a time, and summing the difference from the
result with no error correction provides a large correction to the energy.
This correction greatly reduces the coherence requirements to obtain chemical
accuracy; error rates can be up to two orders of magnitude larger, without the
need for full error correction. This is at
a relatively low overhead; for example, just an additional $\mathcal{O}(n_{qubits})$
evaluations on the quantum computer with only single qubit error correction.
This correction relies on cancellation of error between
  expectation values; each expectation value calculated needs a sufficient
  number of measurements to allow for this cancellation.
On future quantum devices, either no error correction (or possibly limited error
correction) on all qubits and sweeping through full error correction on each
qubit can allow much larger systems to be 
computed, compared to full error correction on each qubit. Though we show
results only for VQE, the method can reduce the error in any measured
observable, and thus has application to a wide range of quantum algorithms. 
Further study on other algorithms, such as quantum phase
estimation~\cite{kitaev-arxiv-1995} and quantum approximate optimization
algorithm~\cite{farhi2014quantum}, should be done to understand 
the impact of this scheme on other algorithms. The magnitude of the correction can also be
used as a metric for measuring how close to the true answer one is, without
knowing the true answer; as the
correction gets smaller, the effect of the environmental noise is smaller.
Furthermore, the magnitude of the correction can be used to compare different
architectures, without need for knowledge of the true answer. Whichever
architecture gives a smaller correction is likely closer to the true answer, and
can be considered a better architecture for that problem.
\begin{acknowledgments}
This work was performed at the Center for Nanoscale Materials, a U.S. Department
of Energy Office of Science User Facility, and supported by the U.S. Department
of Energy, Office of Science, under Contract No. DE-AC02-06CH11357. We
gratefully acknowledge the computing resources provided on Bebop,
a high-performance computing cluster operated by the Laboratory Computing
Resource Center at Argonne National Laboratory. 
\end{acknowledgments}
\section{Appendix}
\appendix*
\section{Derivation of the Observable Correction Formula}
First consider the Lindbland master equation, Eq.~\eqref{master_equation}.
A formal solution for a density matrix evolving from time 
$t$ to $t+\tau$ and satisfying this equation
is
\begin{equation}
\rho (t+\tau ) =   V_{\tau}(\rho (t)), 
\end{equation}
where 
\begin{equation}
V_{\tau}()= \exp [\tau L() ].
\end{equation}
We  use $()$ above to indicate that $L$ and $V$ are superoperators
that take in an operator with the brackets to generate a new one.  
To first order in $\tau$ and taking $L()$ to be the sum over Lindblad
operators of Eq.~\eqref{master_equation},
\begin{equation}\label{approx}
V_{\tau}() \approx 1 + \tau \sum_{i=1}^m L_i(),
\end{equation}

The sequence:  apply gate 1, evolve under $V_{\tau}$, 
apply gate 2, etc., up to gate G corresponds 
exactly to a final density matrix given by
\begin{multline}\label{mixed}
\rho (T) = U_G V_{\tau} \bigg( ~ U_{G-1} \cdots ~
V_{\tau} \big( ~ U_2
(~ V_{\tau}(~U_1 \rho (0) U_1^\dagger ~ ) 
U_2^\dagger ~ \big)
\cdots \\ 
U_{G-1}^\dagger ~ \bigg)  U_G^\dagger.
\end{multline}
Notice that 
Eq.~\eqref{mixed} is not symmetric, with the $V_{\tau}$ operators always 
acting on the right side.

Equation~\eqref{mixed} is $\rho (T)$ for the case of all 
$m$ error sources present.  
The corresponding $\rho_a(T)$ for perfect error correction or no error
terms present
is simply:
\begin{equation}\label{rhoa}
\rho_a (T) = U_G  U_{G-1} \cdots ~
U_2
 U_1 \rho (0) U_1^\dagger  
U_2^\dagger
\cdots U_{G-1}^\dagger  U_G^\dagger.
\end{equation}
The density matrices resulting from removing 
error sources $i$ = 1,2,..., $m$ separately are
\begin{multline}\label{mixed-i}
\rho_i (T) = U_G V_{\tau}^i \bigg( ~ U_{G-1} \cdots ~
V_{\tau}^i \big( ~ U_2
(~ V_{\tau}^i(~U_1 \rho (0) U_1^\dagger ~ ) 
U_2^\dagger ~ \big) 
\cdots \\ 
U_{G-1}^\dagger ~ \bigg)  U_G^\dagger.
\end{multline}
$V_{\tau}^i$ is the corresponding Lindblad evolution operator
that does not contain $L_i$ but has all other terms,
e.g. to first order in $\tau$,
\begin{equation}\label{approx-i}
V_{\tau}^i() \approx 1 + \tau \sum_{j \neq i }^m L_j(),
\end{equation}
Now we consider the new density matrix defined as
\begin{equation}\label{rho-correct}
\tilde{\rho} (T) 
= \rho (T) - \sum_{i=1}^m (\rho(T) - \rho_i (T)).
\end{equation}
Insertion of the Eqs.~\eqref{approx} and \eqref{approx-i} 
into Eq. \eqref{rho-correct} leads, after some tedious but straightforward
algebraic
manipulations, to 
\begin{equation}
\tilde{\rho} (T) \approx \rho_a (T)  
\end{equation}
being correct to first order in $\tau$, i.e. all first order error terms 
exactly cancel,
with  remaining error terms on the order of $\tau^2$ and higher.   
In contrast, the
uncorrected density matrix $\rho (T)$ contains first-order error terms. 

Equation~\eqref{rho-correct} immediately leads to the observable correction
formula of the text, Eq.~\eqref{correction}, when traced
with the observable, A: 
\begin{align}
\begin{split}
\tilde{A} &= \Tr (\tilde{\rho} (T) A) \\ 
        &= \Tr (\rho (T) A) - \sum_{i=1}^m [\Tr (\rho (T)A) - \Tr (\rho_i (T) A)  ]  \\
        &= \langle A \rangle - \sum_{i=1}^m ( \langle A \rangle -\langle A_i \rangle),
\end{split}
\end{align}
which will also be accurate to order $\tau$ whereas the uncorrected observable,
$\langle A \rangle$,
has first order error terms.

\section{Entangling Ansatz for H$_2$}

\begin{figure}
  \centering
  \includegraphics[width=0.9\columnwidth]{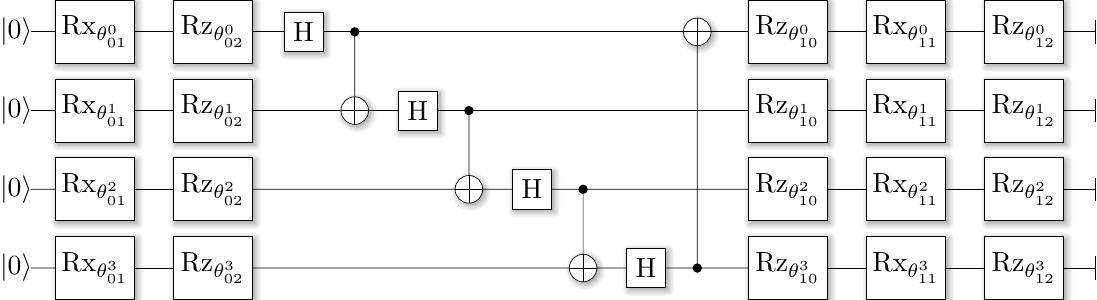}
  \caption{
    Single layer entangling ansatz example for four qubits.
  } 
\label{entangling_circuit}
\end{figure}

\begin{figure}
  \centering
  \includegraphics[width=0.9\columnwidth]{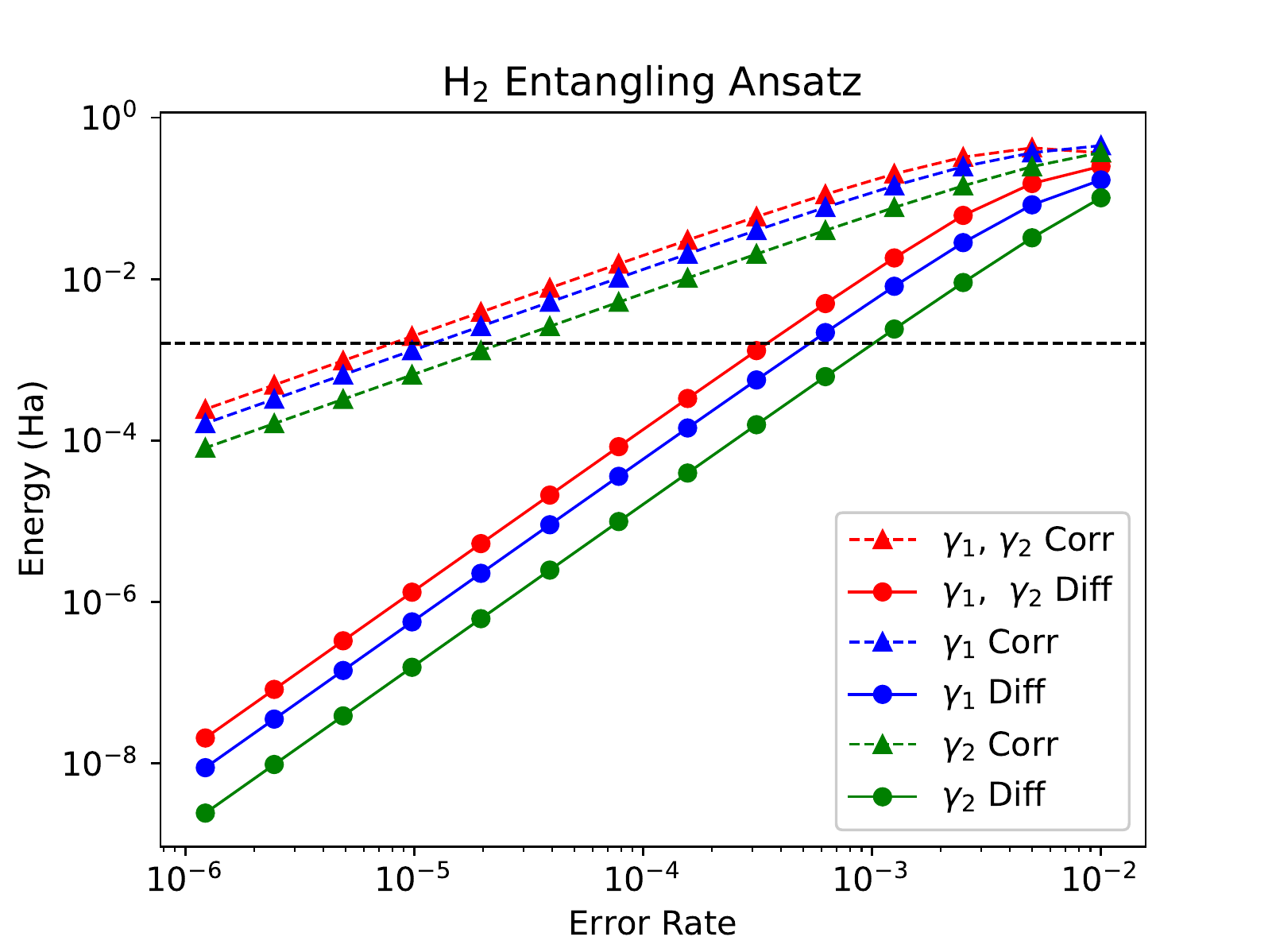}
  \caption{
    Error cancellation for H$_2$ at bond length 0.74 \AA~for an `entangling'
    ansatz with 4 layers (described in the appendix), under amplitude damping ($\gamma_1$), dephasing
    ($\gamma_2$), and both amplitude damping and dephasing ($\gamma_1, \gamma_2$)
    noise sources. The horizontal line represents `chemical
    accuracy', 1.6 mHa. The dashed lines with triangle markers represent the
    amount of correction applied by our scheme. The solid lines with circle
    markers represent the difference between the corrected energy and the energy
    evaluated with no noise. The results are nearly the same as the unitary
    coupled cluster singles doubles ansatz, with corrections of up to 70 mHa applied to get to
    chemical accuracy and error rates which can be up to 50$\times$ larger.    
  } 
\label{h2_example_ibm}
\end{figure}

We also applied our correction scheme to a different wavefunction ansatz, which
 we call the entangling ansatz, similar to that described in~\cite{kandala-nature-2017}.
For
the entangling ansatz, we parameterize the wavefunction by first applying
parameterized rotations
$Rx(\theta^q_{01}) Rz(\theta^q_{02})$, and then alternating between entangling
all the qubits in ring by applying a Hadamard gate 
to qubit $q$ and then a CNOT gate between qubits $q$ and $(q+1) \% n_{qubits}$
for all qubits $q$, where $\%$ is the modulo operator,
followed by parameterized rotations $Rz(\theta^q_{i0})$,
$Rx(\theta^q_{i1})$, and $Rz(\theta^q_{i2})$ for each qubit $q$ and layer $i$.
This is done for a number of layers $d$. This circuit is equivalent to that
described in Ref.~\cite{kandala-nature-2017}, except the entangling circuit is
replaced with the ring of Hadamard - CNOTs. Figure~\ref{entangling_circuit}
gives an example of this circuit with $d=1$. The results are shown in
figure~\ref{h2_example_ibm} for a circuit of $d=4$, and are almost
indistinguishable from the UCCSD results. Corrections of up to 70 mHa are
applied to get to chemical accuracy; error rates can be 50$\times$ larger to get
to chemical accuracy.

\bibliography{vqe_ec}
\end{document}